\documentclass[a4paper,11pt]{article}
\usepackage{amsfonts,latexsym,amssymb}
\input amssymb.sty
\usepackage{epsfig}

\usepackage[applemac]{inputenc}
\usepackage{amsfonts}

\advance\hoffset by -25mm
\def \Z{{\mathbb  Z}}
\def \N{{\mathbb  N}}

\def \R{{\mathbb R}}
\def\E{{\mathbb E}}
\def\B{{\mathbb B}}
\def\I{{\mathbb I}}
\def\ep{\varepsilon}
\renewcommand{\le}{\leqslant}
\renewcommand{\ge}{\geqslant}

\renewcommand{\phi}{\varphi}

\newtheorem{lemma}{Lemma}
\newtheorem{theorem}{Theorem}
\newtheorem{corollary}{Corollary}

\newtheorem{definition}{Definition}
\newtheorem{remark}{Remark}

\newtheorem{proposition}{Proposition}

\begin{document}
\title{\bf Stochastic gene expression\\
    in switching environments}
\author{ Martin Gander, Christian Mazza\thanks{
Section de Math\'ematiques, 2-4 Rue du 
Li\`evre, CP64, CH-1211 Gen\`eve 4, martin.gander@math.unige.ch, christian.mazza@math.unige.ch}
 and Hansklaus Rummler\thanks{D\'epartement de
Math\'ematiques, Chemin du Mus\'ee 23, CH-1700 Fribourg, 
hansklaus.rummler@unifr.ch}}
\date{ February 2, 2005}

\maketitle

\setcounter{page}{0}

\thispagestyle{empty}

\begin{abstract}

We study a stochastic model proposed recently in the genetic literature
to explain the heterogeneity of cell populations or of gene products.
Cells are located in two colonies, whose sizes  fluctuate as
birth with migration processes in  switching environment.
We prove that there is a range of parameters where
heterogeneity induces a larger mean fitness.
\end{abstract}


\vfill\vfill

\vfill

\pagebreak


\section{Introduction}\label{s.results}

In \cite{Thattai}, the authors introduce a model for stochastic gene
expression to study the heterogeneity of cell populations. They assume
that the cells, or for example the product of some gene, can be in two
distinct states or colonies. Let $X(t)$ and $Y(t)$ be the sizes of
these colonies, which are here considered as birth with migration
processes. We assume that the birth rates are either $\gamma_1$ or
$\gamma_0$ with $\triangle\gamma=\gamma_1-\gamma_0>0$, and that the
associated migration rates $k_1$ and $k_0$ are such that $k_0\ge k_1$,
that is cells located in the colony having the smaller birth rate
migrate at a higher rate to the colony with the higher birth rate than
the other way round.

If the birth and migration rates are assigned once and for all to a
corresponding colony (e.g. $\gamma_0$ and $k_0$ to $X$, and $\gamma_1$
and $k_1$ to $Y$), then the mean sizes $n_0(t)=\E(X(t))$ and
$n_1(t)=\E(Y(t))$ satisfy the pair of differential equations (see
\cite{Renshaw73} or \cite{Renshaw91})
\begin{equation}\label{eqn0}
  \begin{array}{rcl}
    \displaystyle
    \frac{{\rm d}n_0(t)}{{\rm d}t} & = & (\gamma_0-k_0)n_0(t)+k_1 n_1(t),\\
    \displaystyle
    \frac{{\rm d}n_1(t)}{{\rm d}t} & = & (\gamma_1-k_1)n_1(t)+k_0 n_0(t).
  \end{array}
\end{equation}
According to \cite{Thattai}, we say that cells of the first colony
represented by $X(t)$ are {\it unfit} (they have the lower birth
rates), and conversely that cells of the second colony represented by
$Y(t)$ are {\it fit}. The proportion of fit cells in the global
population, $y(t)=n_1(t)/(n_1(t)+n_0(t))$, $t\ge 0$, satisfies the
non-linear differential equation
\begin{equation}\label{Fit}
  \frac{{\rm d}y(t)}{{\rm d}t}=
    k_0+(\triangle\gamma-k_0-k_1)y(t)-\triangle\gamma y(t)^2.
\end{equation}

Then, as $t\to\infty$, $y(t)\longrightarrow y_1$, where
$\frac{1}{2}\le y_1\le 1$ follows directly from (\ref{Fit}), see
Section \ref{s.Stationary}. This describes the equilibrium value of
the proportion of fit cells in a non-changing environment. Fixing the
values of the parameters $k_0$, $\gamma_0$ and $\gamma_1$, we can ask
for the value of $0\le k_1\le k_0$ which maximizes the proportion of
fit cells, i.e. the equilibrium value of $y(t)$: the optimal strategy
is to keep all the fit cells in the fit state, that is to set their
migration rate to zero, $k_1=0$. This leads to $y_1=1$, and
thus the optimal solution would be a homogeneous population.

Observations reveal however that most cell populations are not
homogeneous; to explain this, the authors of \cite{Thattai} propose to
introduce a small modification in the model by allowing environmental
changes (for related questions in this context, see
e.g. \cite{Renshaw91}), and show through Monte-Carlo simulations that
the homogeneous solution $k_1=0$ is then not always optimal. The idea
in their model is to allow the birth and migration rates to switch at
random times from one colony to the other, so that cells in the fit
colony become unfit and vice versa.  If for example an environmental
change occurs at some random time $T_1>0$ ($T_0=0$), then the function
$f_1(t)$ representing the proportion of fit cells solves (\ref{Fit})
up to time $T_1$, and just after $T_1$, say at time $T_1+0$, the fit
cells corresponding to $Y(t)$ become unfit and vice versa.  The
proportion of fit cells $f_1(t)$ is then switched to
$f_1(T_1+0)=1-f_1(T_1)$. After $T_1$, the random process
$\{f_1(t)\}_{t\ge 0}$ solves (\ref{Fit}) with initial data
$f_1(T_1+0)$ at time $T_1+0$, until a new environmental change occurs,
say at time $T_2>T_1$.  There is a new switch, and the process is
again solution of (\ref{Fit}), until a new event occurs and so on.

In \cite{Thattai}, the fluctuations of the environment are modeled
using a renewal process; the instants $T_i$, $i\ge 0$, are such that
the sequence of random variables $\{t_i\}_{i\ge 1}$ given by
$t_i\equiv T_i-T_{i-1}$, $i\ge 1$, is i.i.d. distributed according to
some law $\mu$ on $\R^+$. The authors then use Monte-Carlo simulations
to estimate the limiting value of the time averages along trajectories
of the process $f_1(t)$, of the form
$$
  S_N=\frac{1}{T_N}\int_0^{T_{N}}f_1(s)\,{\rm d}s.
$$
This limiting average value is denoted by ${\rm Av}(f_1)_{k_1}$ to
express its dependency on the migration rate $k_1<k_0$, when all the
remaining parameters are fixed.  Their simulations indicate that there
is a range of parameters ($k_0$ not too large) such that
$$
  {\rm Av}(f_1)_{k_1>0}>{\rm Av}(f_1)_{k_1=0},
$$
which means that heterogeneous populations are more adapted than
homogeneous ones in a switching environment.  

In this paper, we study mathematically the limiting behavior of the
stochastic process $f_1(t)$ and the associated time average $S_N$ by
giving its stationary measure, and we provide mathematical formulas and
numerical solutions, which might be of interest in practical
laboratory experiments (see e.g. \cite{Thattai}).
 
Our technique uses the process $X_k=f_1(T_k-0)$, $X_0=f_1(0)$, which
is such that $X_{k+1}=\phi_{t_{k+1}}(1-X_k)$, for some mapping
$\phi_t(x)$ (see Definition \ref{MarkovChain}).  $(X_k)_{k\ge 0}$ is a
stochastic recursive Markov chain, and $S_N$ can be expressed as an
additive functional of the trajectory of $(X_k)_{1\le k\le N}$.  In
Section \ref{s.Stationary}, we recall a Theorem from \cite{Diaconis}
on the convergence of stochastic recursive chains, which applies in
this setting. We give conditions ensuring the existence and uniqueness
of a stationary measure $\pi$, as well as geometric ergodicity.  In
Section \ref{Stationarity}, we consider the case where $\mu$ is
exponential of parameter $\kappa >0$, and show that $\pi$ has a ${\cal
C}^{\infty}$ density $P$ with respect to Lebesgue measure. We
furthermore prove in Theorem \ref{ergodic} that a multiple $G$ of $P$
solves a second order differential equation with weak
singularities. Proposition \ref{DiffSolutions} provides series
expansions for $P$, which are necessary to derive properties of $P$
near the singularities. In Section \ref{s.numerical}, we show
numerical solutions, using the series expansions of Proposition
\ref{DiffSolutions} to start the numerical integration. We provide
an example where ${\rm Av}(f_1)_{k_1>0}>{\rm
Av}(f_1)_{k_1=0}$, which shows that it can be better to allow fit
cells to migrate to the unfit state than to conserve all the fit cells
in the fit state in such a switching environment. This is a regime
where it is suitable for the colonies to anticipate bad hypothetical
future events.

\section{Convergence of recursive chains}\label{s.Stationary}

We first give some basic results for the differential equation
(\ref{Fit}). The right hand side of (\ref{Fit}) can be factored into
$-\triangle\gamma (y-y_0)(y-y_1)$, where
$y_0=(\triangle\gamma-k_0-k_1)-\sqrt{d})/(2\triangle\gamma)<0$,
$y_1=((\triangle\gamma-k_0-k_1)+\sqrt{d})/(2\triangle\gamma)>0$, and
$d=(\triangle\gamma-k_0-k_1)^2+4k_0\triangle\gamma$.  Then $k_0>k_1$
implies that $ 0<1-y_1<\frac{1}{2}<y_1<1, $ and that the derivative
${\rm d}f_1(t)/{\rm d}t$ is positive when $f_1(t)$ is in the interval
$[0,y_1)$, negative in $(y_1,1]$, and it vanishes for $f_1(t)=y_1$.
It is not hard to check that any realization of the trajectory
$\{f_1(t)\}_{t\ge 0}$, with initial data $f_1(0)\in I=(1-y_1,y_1)$
will remain forever in $I$, and that any trajectory starting in the
interval $I^c=[0,1]\setminus I$ will enter $I$ after an almost surely
finite time. (However, $f_1(0)=y_1$ implies $f_1(t)\equiv y_1$.) We
thus restrict our study to the interval $I$.

Given $t\in\R^+$, we define the mapping $\phi_t:I\longrightarrow I$
such that $\phi_t(x)$ is the value of the solution of (\ref{Fit}) at
time $t$ when starting at $x\in I$ at time $t_0=0$.  Using separation
of variables for (\ref{Fit}), we obtain
the relation
\begin{equation}\label{BasicRelation}
  \frac{\phi_t(x)-y_0}{y_1-\phi_t(x)}=\frac{x-y_0}{y_1-x}\exp(\beta t),
\end{equation}
where we set $\beta=\triangle\gamma (y_1-y_0)$.  Given $u\in I$, let
$\delta t(u,y)$ denote the time interval the orbit of the dynamical
system (\ref{Fit}) needs to join $u$ and $y$, $y\ge u$, when starting
at time $t=0$ at $u$. Then
\begin{equation}\label{TimeInterval}
  \beta\delta t(u,y)=\ln\Big(\frac{(y-y_0)(y_1-u)}{(y_1-y)(u-y_0)}\Big).
\end{equation}

\begin{definition}\label{MarkovChain}
  Given $X_0=f_1(0)\in I$, consider the Markov chain with values in $I$
  defined by
  $$
    X_{k+1}=\phi_{t_{k+1}}(1-X_k),
  $$
  where the sequence of random variables $\{t_k\}_{k\in\N^+}$ is
  i.i.d. distributed according to some law $\mu$ on $\R^+$. This
  Markov chain describes the evolution of $f_1(T_k-0)$, at the
  instants just before the switches, with $T_{k+1}-T_k=t_{k+1}$.
\end{definition}
We first recall and adapt results of \cite{Diaconis} on the
convergence of such Markov chains, also called {\it stochastic
recursive chains}, see e.g. \cite{Borovkov}.  The general setting is
described by a complete separable metric space $(S,\rho)$, the set of
values taken by the Markov chain, a family of mappings $f_{\theta}:\
S\longrightarrow S$, indexed by parameters $\theta$ living in some
parameter space ${\Theta}$, and a probability measure $\mu$ on
${\Theta}$. Given an i.i.d.\ sequence of random elements $\theta_n$,
$n\ge 1$, of law $\mu$, we can consider the Markov chain
$(X_n)_{n\in\N}$ given by $X_{n+1}=f_{\theta_{n+1}}(X_n)$.  The
following Theorem gives conditions for the existence and uniqueness of
a stationary measure (Theorem 1.1 of \cite{Diaconis}). In what
follows, $P^{(n)}(x,{\rm d}y)$ denotes the law of the Markov chain
$X_n$ and $\rho[P^{(n)}(x,\cdot),\pi]$ is the Prokhorov metric, see
below.
\begin{theorem}\label{DiacConv}
Assume that the family of functions $f_{\theta}$, $\theta\in \Theta$
is Lipschitz with
$$
  \rho(f_{\theta}(x),f_{\theta}(y))\le K_{\theta}\rho(x,y),\ \ x,y\in
  S,
$$
$\forall \theta\in \Theta$. Assume furthermore that
\begin{equation}\label{TrivCondi}
  \int K_{\theta} \mu({\rm d}\theta)<\infty,\ \ \int
  \rho(f_{\theta}(x_0),x_0)\mu({{\rm d}\theta})<\infty,
\end{equation}
for some $x_0\in S$, and that
\begin{equation}\label{Contraction}
  \int \ln(K_{\theta})\mu({{\rm d}\theta})<0.
\end{equation}
Then
\begin{itemize}
  \item The Markov chain has a unique stationary distribution $\pi$,
  \item $\rho[P^{(n)}(x,\cdot),\pi]\le A_x r^n$, for constants $A_x$
    and $r$ with $0<A_x<\infty$ and $0<r<1$; this bound holds for all
    times $n$ and all starting positions $x$, 
  \item the constant $r$ does not depend on $n$ or $x$; the constant
    $A_x$ does not depend on $n$, and $A_x<a+b\rho(x,x_0)$, where
    $0<a,b<\infty$.
\end{itemize}
\end{theorem}
In our setting, $S$ is given by $I$ and the parameter set
$\Theta$ is just $\R^+$.  The Prokhorov distance
$d_n:=\rho[P^{(n)}(X_0,\cdot),\pi]$ is the infimum of the $\delta >0$
such that
\begin{equation}\label{Prok}
  P^{(n)}(X_0,C)<\pi(C_{\delta})+\delta\quad 
    \mbox{and}\quad \pi(C)<P^{(n)}(X_0,C_{\delta})+\delta,
\end{equation}
where $C$ runs over the Borel sets of $I$ and, for given
$C\in\B(I)$, $C_{\delta}$ denotes the set of points of $I$
whose distance from $C$ is less than $\delta$ (see Section 5.1 of
\cite{Diaconis}). Condition (\ref{Contraction}) means that the
functions $f_{\theta}$ are contractions in the average.  We first
express this condition in our setting: for $t\in \Theta=\R^+$ and
$u\in I=S$, the mapping $\phi_t(u)$ is given explicitly by
\begin{equation}\label{ExplicitPhi}
  \phi_t(u)=\frac{y_0(y_1-u)+y_1(u-y_0)\exp(\beta t)}{y_1-u
    +(u-y_0)\exp(\beta t)}.
\end{equation}
 Setting $f_t(x)=\phi_t(1-x)$, we obtain
\begin{lemma}\label{Lip}
For all $t\in\R^+$,
$$
  \frac{{\rm d}}{{\rm d}x}f_t(x)=-\frac{(y_1-y_0)^2\exp(\beta t)}
    {(y_1-1+x+(1-x-y_0)\exp(\beta t))^2},
$$
$$
  K_t := \sup_{x\in I}\vert \frac{{\rm d}}{{\rm d}x}f_t(x)\vert
    =\frac{(y_1-y_0)^2\exp(\beta t)}{(2y_1-1+(1-y_1-y_0)\exp(\beta
    t))^2}.
$$
If $\mu$ is exponential of parameter $\kappa >0$, and
$\alpha=\kappa/\beta$, then the conditions given in (\ref{TrivCondi})
are satisfied, and
$$
  \int_{\R^+}\kappa \exp(-\kappa t)\ln(K_t) {\rm d}t=
   -\alpha-2z \int_0^{\infty}\frac{\exp(-(1+\alpha)t)}
   {1-z \exp(-t)}{\rm d}t,
$$ 
where we set $z=-(2y_1-1)/(1-y_1-y_0)<0$. Condition
(\ref{Contraction}) is thus satisfied if
\begin{equation}\label{Condi}
  -\alpha-2z \int_0^{\infty}\frac{\exp(-(1+\alpha)t)}{1-z \exp(-t)}{\rm d}t<0.
\end{equation}
\end{lemma}
\begin{remark}
  When $\vert z\vert\le 1$, the integral
  $\int_0^{\infty}(\exp(-(1+\alpha)t))/(1-z \exp(-t)){\rm d}t$ is the
  Lerch Phi function $\Phi(z,s,v)=\sum_{n\ge 0}(v+n)^{-s}z^n$, with
  $s=1$ and $v=1+\alpha$, and is also equal to Gauss's Hypergeometric
  function $_2{\rm F}_1(1,1+\alpha;2+\alpha;z)/(1+\alpha)$ (see e.g.
  \cite{Erdelyi}, chap. 1.11).
\end{remark}
{\bf Proof}:
 Taking the derivative of (\ref{ExplicitPhi}) with
   respect to $u$, we obtain
   $$
    \frac{{\rm d}}{{\rm d}u}\phi_t(u)=
    \frac{(y_1-y_0)^2\exp(\beta t)}{\bigl(y_1-u+(u-y_0)\exp(\beta t)\bigr)^2}
   $$
     and thus
   $$
      f'_t(x)=-\frac{\rm d}{{\rm 
d}u}\phi_t(1-x)=-\frac{(y_1-y_0)^2\exp(\beta t)}
      {(y_1-1+x+(1-x-y_0)\exp(\beta t))^2}<0,
   $$
   as required. The expression for $K_t$ follows from direct
   computation.

\section{Convergence to stationarity in Poissonian environments}
  \label{Stationarity}

Assume that $\mu$ is exponential of parameter $\kappa >0$.  We will
see in the sequel that the stationary measure $\pi$ has, under some
conditions, a density $P(y)$ such that with
$Q(y)=((y-y_0)/(y_1-y))^{\alpha}$, where $\alpha=\kappa/\beta$, the
function $G(y)=P(y)Q(y)(y-y_0)(y_1-y)$ satisfies the differential
equation
\begin{equation}\label{Diff}
  G''(y)+ U(y)G'(y)+ V(y)G(y)=0,
\end{equation}
where
$\tilde y_0=1-y_0$, $\tilde y_1 =1-y_1$,
\begin{equation}\label{Conda}
U(y)=\frac{\alpha+1}{y-\tilde
y_0}-\frac{\alpha-1}{y-\tilde 
y_1}+\frac{\alpha}{y-y_1}-\frac{\alpha}{y-y_0},
\end{equation}
and
\begin{equation}\label{Condb}
V(y)=\frac{\alpha^2
(y_1-y_0)^2}{(y-y_0)(y-y_1)(y-\tilde y_0)(y-\tilde y_1)}.
\end{equation}
The following proposition will therefore be useful:
\begin{proposition}\label{DiffSolutions}
  The solutions of the second order homogeneous linear differential
  equation (\ref{Diff}) are analytic on the interval
  $I=(\tilde y_1,y_1)$. Two fundamental solutions $\tilde
  G_1(y)$, $\tilde G_2(y)$ are
\begin{itemize}
   \item $\tilde G_1(y)=(y-\tilde y_1)^\alpha \tilde W_1(y)$, where
     $\tilde W_1(y)$ is analytic on $(\tilde y_1-\delta,y_1)$ for some
     $\delta>0$ and with $\tilde W_1(\tilde y_1)=1$.

   \item $\displaystyle \tilde G_2(y)=\left\{
     \begin{array}{ll}
     \tilde W_2(y),& {\rm if}\ \alpha \not\in\Z,\\
     \tilde W_2(y)
     +\tilde C\tilde G_1(y)\ln(y-\tilde y_1),& {\rm if}\ \alpha\in\Z,
     \end{array}
     \right.$\\ 
     with $\tilde W_2(y)$ analytic on $(\tilde
     y_1-\delta,y_1)$ for some $\delta>0$, $\tilde W_2(\tilde y_1)=1$
     and $\tilde C\in\R$.
\end{itemize}
Another set of two fundamental solutions $G_1(y)$, $G_2(y)$ is
\begin{itemize}
  \item $G_1(y)=(y_1-y)^{1-\alpha}W_1(y)$, where $ W_1(y)$ is analytic
    on $(\tilde y_1,y_1+\delta)$ for some $\delta>0$ and with
    $W_1(y_1)=1$.

  \item
     $\displaystyle G_2(y)=\left\{
     \begin{array}{ll}
     W_2(y),& {\rm if}\ \alpha \not\in\Z,\\
     W_2(y)
     +CG_1(y)\ln(y_1-y),& {\rm if}\ \alpha\in\Z,
     \end{array}
     \right.$\\ 
     with $W_2(y)$ analytic on $(\tilde y_1,y_1+\delta)$
     for some $\delta>0$, $W_2(y_1)=1$ and $C\in\R$.
\end{itemize}

\end{proposition}
In the appendix, we show this result for completeness, and also how
these fundamental solutions can be computed by series expansion about
$\tilde y_1$ and $y_1$ respectively.

\begin{theorem}\label{ergodic}
Assume that
$$
  -\alpha-2z \int_0^{\infty}\frac{\exp(-(1+\alpha)t)}
  {1-z \exp(-t)}{\rm d}t <0,
$$ 
where $z=-(2y_1-1)/(1-y_1-y_0)<0$. Then the Markov chain $X_k$ from
Definition \ref{MarkovChain}, with initial data $X_0\in I=(1-y_1,y_1)$
has a unique stationary distribution $\pi$ of ${\cal C}^{\infty}$
density
$$
  P(y)=\frac{Q(y)^{-1}(y-\tilde y_1)^{\alpha}
    \tilde W_1(y)/(y_1-y)/(y-y_0)}{\int_I
    Q(z)^{-1}(z-\tilde y_1)^{\alpha}
    \tilde W_1(z)/(y_1-y)/(y-y_0)\,{\rm d}z}.
$$
Here, $Q(y)=\Big(\frac{y-y_0}{y_1-y}\Big)^{\alpha}$, where
$\alpha=\kappa/\beta$, $\tilde W_1(y)$ is the analytic function on
$(\tilde y_1-\delta,y_1)$ with $\tilde W(\tilde y_1)=1$, such
that $\tilde G_1(y)=(y-\tilde y_1)^\alpha \tilde W_1(y)$ is a
solution of the differential equation (\ref{Diff}). In the
neighborhood of $y=y_1$, this solution is such that $0<\lim_{y\to
y_1}\tilde W_1(y)<+\infty$. Finally, the behavior of the density $P$ near
$y_1$ is given by $(y_1-y)^{\alpha -1}$, and thus converges when
$\alpha\ge 1$ and diverges toward $+\infty$ when $\alpha <1$.  Let
$f(x)=x$ and $g(x)=\ln((x-1+y_1)/(y_1-x))$ be defined on $I$.  Then $
g \in {\rm L}^1(I,\B(I),\pi)$ with
\begin{equation}\label{Moment}
  \E_{\pi}(f)=y_0+\frac{\kappa}{\triangle\gamma}\E_{\pi}(g).
\end{equation}
\end{theorem}
\begin{remark} 
  Relation (\ref{Moment}) will be useful
  when considering time averages for Monte-Carlo simulations, see 
  Section \ref{Average}.
\end{remark}
{\bf Proof}:
The existence and uniqueness of the stationary measure follows
from Theorem \ref{DiacConv} and Lemma \ref{Lip}.
Let $Y$ be a random variable of law $\pi$, and
let $T$ be exponential of parameter $\kappa >0$, independent
of $Y$. In the stationary regime,
$Y=_{{\cal L}} \phi_T(1-Y)$.
Let $F(y)=P(Y<y)$. Then
$$
  F(y)=\int_{I\times\R^+}\pi({\rm d}v)\kappa \exp(-\kappa t)
    \I(\phi_t(1-v)<y) {\rm d}t,
$$
where $\I(\cdot)$ denotes the indicator function. For given $y\in I$,
the time variable $t$ is restricted to the interval $0\le t< \delta
t(\tilde y_1,y)$ , see (\ref{TimeInterval}). Thus
$$
  F(y)=\int_{0}^{\delta t(1-y_1,y)}\kappa \exp(-\kappa t)\int_I\pi({\rm d}v)
    \I(\phi_t(1-v)<y) {\rm d}t .
$$ 
For given $t$ in this interval, the set of $v\in I$ with $\phi_t(1-v)<
y$ is given by
$$
  \{v\in I;\ 1-v<\frac{y_1(y-y_0)+\exp(\beta t)(y_1-y)y_0}
   {y-y_0+\exp(\beta t)(y_1-y)}\}.
$$ 
It follows that $\int_I\pi({\rm d}v)\I(\phi_t(1-v)<y)=1-F(1-u)$, where
we set $u=\bigl(y_1(y-y_0)+\exp(\beta t)(y_1-y)y_0\bigr)/
\bigl(y-y_0+\exp(\beta t)(y_1-y)\bigr)$, with $t=\delta t(u,y)$. We
make the change of variable $t=\delta t (u,y)$ with
$$
  \frac{{\rm d}t}{{\rm d}u}=-\frac{y_1-y_0}{\beta(y_1-u)(u-y_0)}.
$$
Then
$$
F(y)= \alpha\Big(\frac{y_1-y}{y-y_0}\Big)^{\alpha}\int_{1-y_1}^y
\frac{y_1-y_0}{(y_1-u)(u-y_0)}
\Big(\frac{u-y_0}{y_1-u}\Big)^{\alpha}(1-F(1-u)){\rm d}u.
$$
This is a fixed point equation for the distribution function $F$. We
use it for proving that the probability measure $\pi$ has a ${\cal
C}^{\infty}$ density on the interval $I$. First notice that $F$ is
monotonically increasing and integrable on $I$. The above relation
then shows that $F$ is continuous on $I$. Using again this argument
recursively, one sees that $F$ is the integral of a continuous function
and is therefore differentiable, with a continuous derivative. The
${\cal C}^{\infty}$ differentiability is obtained by iterating this
argument. Let $P$ be the ${\cal C}^{\infty}$ density of $\pi$ with
respect to Lebesgue measure. Our strategy runs as follows: We use the
fixed point relation to show that a multiple $G$ of $P$ satisfies a
second order differential equation, which has only weak singularities,
and then deduce properties of $P$ with the help of Proposition
\ref{DiffSolutions}.

For given $v\in I$, the time variable $t$ is restricted to the
interval
$$
  0\le t\le \delta t(u,y)=
   \ln\bigl((y-y_0)(y_1-y)/(y_1-y)(u-y_0)\bigr)/\beta,
$$
where $u=1-v$ (see \ref{TimeInterval}).  It follows that
$$
  F(y)=\int_{1-y}^{y_1}P(v){\rm d}v\int_{0}^{\delta t(u,y)}\kappa
   \exp(-\kappa t){\rm d}t,
$$
with
\begin{eqnarray*}
  P(y)=\frac{{\rm d}F(y)}{{\rm d}y}&=&\int_{1-y}^{y_1}P(v) {\rm d}v
    \kappa \exp(-\kappa \delta t(u,y))\frac{{\rm d}\delta t(u,y)}{{\rm
    d}y}\\
  &=&\alpha \int_{1-y}^{y_1}{\rm d}v P(v)
    \frac{Q(u)}{Q(y)}\frac{(y_1-y_0)}{(y-y_0)(y_1-y)},
\end{eqnarray*}
where we set $Q(y)=((y-y_0)/(y_1-y))^{\alpha}$.  Using $u=1-v$ and
setting $G(y)=P(y)Q(y)(y-y_0)(y_1-y)$, one gets
\begin{equation}\label{relc}
  G(y)=\int_{1-y_1}^y G(1-u)R(u)H(u)\,{\rm d}u,
\end{equation}
where $R(u)=\alpha Q(u)Q(1-u)^{-1}$ is such that $R(1-u)=\alpha^2/
R(u)$, and $H(u)=(y_1-y_0)/(y_1-1+u)/(1-u-y_0)$.  Taking the
derivative gives
\begin{equation}\label{Deriv}
  G'(y)=G(1-y)R(y)H(y),
\end{equation}
or
$$
  G(1-y)=G'(y)R(y)^{-1}H(y)^{-1}=\alpha^{-2}G'(y) R(1-y)/H(y).
$$
Taking a second derivative then gives
$$
  G''(y)+\frac{{\rm d}}{{\rm d}y}\ln(\frac{R(1-y)}
    {H(y)})G'(y)+\alpha^2H(y)H(1-y) G(y)=0.
$$ 
and simplifying the terms leads to (\ref{Diff}). We see that
$R(u)H(u)\sim (u-1+y_1)^{\alpha -1}$, as $u\to 1-y_1$.   The exponents associated with the
fundamental solutions are $\rho = 0$ or $\alpha$ in the neighborhood
of $y=1-y_1$ and $\rho'=0$ or $1-\alpha$ near $y=y_1$.

Assume first that $\alpha\not\in \N^+$. 
We first check the behavior of $G$ in a neighborhood of
$y=\tilde y_1$. Set $y=\tilde y_1+\ep$, $\ep >0$,
with $1-y=y_1-\ep$.
Proposition \ref{DiffSolutions} shows that $G$
is a linear combination $G(y)=\tilde A \ep^{\alpha}\tilde W_1(y)
+\tilde B \tilde W_2(y)$, for constants $\tilde A$, $\tilde B\in\R$.
Similarly, $G(1-y)=A \ep^{1-\alpha} W_1(1-y)+ B W_2(1-y)$,
for real constants $A$ and $B$. As $\ep\to 0$, the right hand side
of (\ref{relc}) behaves like $\ep^{\alpha}G(y_1-\ep)\to 0$.
Suppose that $\tilde B\ne 0$. Then $G(y)\sim \tilde B \tilde W_2(y)\ne 0$,   
and (\ref{relc}) can't be satisfied. One must thus have $\tilde B =0$,
so that $G(y)=\tilde A \ep^{\alpha}\tilde W_1(y)$. When $\alpha >1$,
(\ref{relc}) implies that $A=0$. It follows that, for arbitrary $\alpha >0$,
$\lim_{y\to y_1}G(y)=B W_2(y_1)\ne 0$, and that 
$G(\tilde y_1+\ep)\sim \tilde A \ep^{\alpha}\tilde W_1(\tilde y_1)$, $\ep\to 0$,
as required. The corresponding result for $P$ follows.

Suppose that $\alpha\in\N^+$. The right hand side of (\ref{relc}) behaves like
$$F(\ep):=\ep^{\alpha}(A \ep^{1-\alpha}W_1(y_1)+B(W_2(y_1)+C \ep^{\alpha-1}W_1(y_1)\ln(\ep))),$$
with $F(\ep)\to 0$ as $\ep\to 0$, and $G(\tilde y_1+\ep)$ behaves like
$$\tilde F(\ep):=\tilde A \ep^{\alpha}\tilde W_1(\tilde y_1)+\tilde B(\tilde W_2(\tilde y_1)+\tilde C \ep^{\alpha}\tilde W_1(\tilde y_1)
\ln(\ep)).$$
One has $\tilde F(\ep)\sim \tilde B \tilde W_2(\tilde y_1)$, $\ep\to 0$, when $\tilde B\ne 0$ and
$\tilde F(\ep)\sim \tilde A \ep^{\alpha}\tilde W_1(\tilde y_1)$, when $\tilde B =0$. 
(\ref{relc}) shows that necessarily $\tilde B =0$. Suppose that $\alpha =1$. Then one must
have $BC=0$, implying the existence of the limit $\lim_{y\to y_1}G(y)\ne 0$. When
$\alpha >1$, $A=0$, $B\ne 0$, and $\lim_{y\to y_1}G(y)=B W_2(y_1)$, as required.

 Finally, we check the
identity (\ref{Moment}).  First $g\in L^1(I,\B(I),\pi)$ follows from
the behavior of the density $P$ at the boundaries of $I$, as described
above. Next,
$$
  \E_{\pi}(g)=\int_I \ln(\frac{y-1+y_1}{y_1-y})P(y){\rm d}y,
$$
where $J:=\int_I\ln(y-1+y_1)P(y){\rm d}y$ is such that
\begin{eqnarray*}
  J&=&\int_I\ln(y-1+y_1)G(y)Q(y)^{-1}\frac{H(1-y)}{y_1-y_0}{\rm d}y\\
  &=&\frac{1}{y_1-y_0}\int_I\ln(y_1-u)G(1-u)Q(1-u)^{-1}H(u){\rm d}u\\
  &=&\frac{1}{\alpha(y_1-y_0)}\int_I
    \frac{\ln(y_1-u)}{Q(u)}G(1-u)R(u)H(u){\rm d}u\\
  &=&\frac{1}{\alpha(y_1-y_0)}\int_I
    \frac{\ln(y_1-u)}{Q(u)}G'(u){\rm d}u\\
  &=&\frac{1}{\alpha(y_1-y_0)}\Big(G(u)
    \frac{\ln(y_1-u)}{Q(u)}\Big\vert_{1-y_1}^{y_1}-\int_IG(u)
    \Big(\frac{\ln(y_1-u)}{Q(u)}\Big)'{\rm d}u\Big)\\
  &=&\frac{1}{\alpha(y_1-y_0)}\int_I\frac{G(u)}{Q(u)}
    \frac{(u-y_0)}{(y_1-u)(u-y_0)}{\rm d}u
    +\int_I\ln(y_1-u)P(u){\rm d}u,
\end{eqnarray*}
where we use (\ref{Deriv}).  It follows that
$$
  \E_{\pi}(g)=\frac{1}{\alpha(y_1-y_0)}
    \E_{\pi}(f)-\frac{y_0}{ \alpha(y_1-y_0)},
$$
proving (\ref{Moment}) since $\alpha=\kappa/(\triangle\gamma(y_1-y_0))$.

\begin{corollary}\label{LimitingLaw}
Assume that condition (\ref{Condi}) holds. Then, as $t\to +\infty$,
the law of the stochastic process $f_1(t)$, $t\ge 0$, $f_1(0)\in I$,
converges toward the stationary measure $\pi$ of density $P$ of the
Markov Chain $X_k$.
\end{corollary}

\noindent {\bf Proof}:
Given $t\in\R^+$, let $t_*$ be the last renewal time before $t$, and
set $S_*=t-t_*$.  When the length of the overlapping random interval
is exponential, $S_*$ is also exponential. In the stationary regime,
or equivalently for large $t$, one has the identity in law
$f_1(t)=_{\cal L} \phi_{S_*}(1-X)$, where $X$ is distributed according
to $\pi$, and the result follows.

\section{Time averages\label{Average}}

When the conclusions of Theorem \ref{DiacConv} hold, the chain $X_k$
has a unique stationary probability measure $\pi$, and $\sum_{k=1}^n
g(X_k)/n$ converges a.s. toward the expectation of $g$ under $\pi$,
for any function $g$ in $L^1(I,\B(I),\pi)$, (see e.g. \cite{Breiman}).
In \cite{Thattai}, the authors use Monte-Carlo methods based on the
process $f_1(t)$, $t\ge 0$, to estimate the mean fitness by
considering the time average
\begin{equation}\label{TimeAverage}
  S_N=\frac{1}{T_{N}}\int_0^{T_{N}}f_1(s)\,{\rm d}s,
\end{equation}
where $N$ is a fixed number of renewal periods.

\begin{lemma}\label{Formula}
Let $N\in\N^+$. Given a realization $0=T_0<T_1<\cdots<T_N$ of the
renewal process, we have
\begin{equation}\label{TimeAverageB}
  \frac{1}{T_N}\int_0^{T_N}f_1(s)\,{\rm d}s =y_0+\frac{(y_1-y_0)}{\beta T_N}
    \ln\Big(\prod_{i=1}^ N\frac{X_{i-1}-(1-y_1)}{y_1-X_i}\Big).
\end{equation}
\end{lemma}

\noindent {\bf Proof}:
Consider the integrals
$$
  \int_{T_{i-1}}^{T_i}f_1(s)\,{\rm d}s,
$$
where $f_1(T_{i-1}+0)=1-X_{i-1}$ and $f_1(T_i-0)=X_i$.  The value of
$y(s)=f_1(T_{i-1}+s)$, $s\in(0,T_i-T_{i-1})$ is given implicitly by
(\ref{BasicRelation}); Therefore
$$
  y(s)=\frac{y_0(y_1-u)+y_1(u-y_0)\exp(\beta s)}
    {y_1-u+(u-y_0)\exp(\beta s)},
$$
where we set $u=1-X_{i-1}$, and thus, after a longer but not difficult
computation, one obtains
$$
  \int_{T_{i-1}}^{T_i}f_1(s)\,{\rm d}s=
    y_0(T_i-T_{i-1})+\frac{y_1-y_0}\beta
    \ln\left(\frac{y_1-u+(u-y_0)\exp\bigl(\beta(T_i-T_{i-1})\bigr)}
    {y_1-y_0}\right),
$$
and the result follows, since
\begin{eqnarray*}
  y_1-u+(u-y_0)\exp(\beta(T_{i}-T_{i-1}))&=&(y_1-u)\bigl(1+\frac{u-y_0}{ 
    y_1-u}\exp(\beta t_i)\bigr)\\
  &=&\frac{(y_1-(1-X_{i-1}))(y_1-y_0)}{y_1-X_i}.
\end{eqnarray*}

\begin{theorem}\label{ConvergenceTimeAverage}
Suppose that $\mu$ is exponential of parameter $\kappa >0$, and assume
(\ref{Condi}).  Let $f(x)=x$ and $g(x)=\ln((x-1+y_1)/(y_1-x))$ be
defined on $I$. Then
$$
  \lim_{N\to\infty}\frac{1}{T_N}\int_0^{T_N}f_1(s)\,{\rm
    d}s=y_0+\frac{\kappa}{\triangle\gamma}
    \E_{\pi}(g)=\E_{\pi}(f),\ \ a.s.
$$
\end{theorem}

\noindent {\bf Proof}:
From equation (\ref{TimeAverageB}), we obtain
$$
  \frac{1}{T_N}\int_0^{T_N}f_1(s)\,{\rm d}s=y_0+\frac{(y_1-y_0)}{\beta
    T_N}\ln(\frac{X_0-1+y_1}{y_1-X_N})
    +\frac{(y_1-y_0)}{\beta T_N}\sum_{i=1}^{N-1}g(X_i).
$$
As $T_N$ is a renewal process with exponential inter arrival times of
parameter $\kappa$, it follows that $T_N/N$ converges a.s. toward
$1/\kappa$. Next, $ g\in L^1(I,\B(I),\pi)$ follows from the behavior
of the density $P$ at the boundaries of $I$, as described in Theorem
\ref{ergodic}. From Proposition \ref{DiffSolutions} and Theorem
\ref{ergodic}, the behavior of $P$ in the neighborhood of $y=1-y_1$ is
given by $(y-1+y_1)^{\rho_1}$ where $\rho_1=\alpha$ and by
$(y_1-y)^{\rho_2+\alpha-1}$ in the neighborhood of $y=y_1$, where
$\rho_2= 0$.  The Markov chain $X_k $ is geometrically ergodic, and
thus the last term converges a.s.  toward
$(\kappa/(\triangle\gamma))\E_{\pi}(g)$. We finally check that
$\ln(y_1-X_N)/N$ converges a.s.  toward 0. Given $\epsilon >0$,
consider the probability
\begin{eqnarray*}
  P(\vert \ln(y_1-X_N)\vert > N\epsilon)&=&P(\ln(y_1-X_N)<-N\epsilon)\\
    &=&P(X_N>y_1-\exp(-N\epsilon))=P^{(N)}(X_0,A_N),
\end{eqnarray*}
where $A_N=\{x>y_1-\exp(-N\epsilon)\}$. Using the behavior of $P$ in
the neighborhood of $y=y_1$, one gets that $\pi(A_N)\le M
(\exp(-\epsilon N))^{\rho_2+\alpha}$, for some positive constant
$M$. Let $\gamma_N:=\vert P^{(N)}(X_0,A_N)-\pi(A_N)\vert$, and let
$d_N$ be the Prokhorov distance defined in (\ref{Prok}). If
$\pi(A_N)\ge P^{(N)}(X_0,A_N)$, then $\gamma_N\le \pi(A_N)$. If
$\pi(A_N)\le P^{(N)}(X_0,A_N)$, one has $P^{(N)}(X_0,A_N)\le
\pi((A_N)_{d_N})+d_N$, and it follows that
\begin{eqnarray*}
  \gamma_N= P^{(N)}(X_0,A_N)-\pi(A_N)&\le&  \pi((A_N)_{d_N})-\pi(A_N)+d_N\\
    &=& \int_{y_1-\exp(-\epsilon N)-d_N}^{y_1-\exp(-\epsilon
      N)}P(y) {\rm d}y +d_N\\
    &\le&d_N +D(\exp(-\epsilon 
      N)^{\rho_2+\alpha}-(d_N+\exp(-\epsilon N))^{\rho_2+\alpha}),
\end{eqnarray*}
for some positive constant $D>0$. Theorem \ref{DiacConv} gives that
$$
  P(\vert \ln(y_1-X_N)\vert >\epsilon N)\le\vert
    P^{(N)}(X_0,A_N)-\pi(A_N)\vert +\pi(A_N)
    \le h(X_0)\lambda^N,
$$
for some bounded function $h$ and a positive number $0<\lambda<1$.
The result then follows from the Borel-Cantelli Lemma. The last
identity is (\ref{Moment}) of Theorem \ref{ergodic}.

\section{Numerical Examples}\label{s.numerical}

We now compute the density $P$ given in Theorem \ref{ergodic}
numerically. To do so, we solve the differential equation (\ref{Diff})
numerically, starting in the neighborhood of the singular point
$y=\tilde y_1=1-y_1$. Proposition \ref{DiffSolutions} and Theorem
\ref{ergodic} show that $\lim_{y\to \tilde y_1}P(y)=0$, and that the
first derivative of $P$ behaves like $(y-\tilde y_1)^{\alpha -1}$,
which goes to $+\infty$ when $\alpha <1$. We start the numerical
solution at the point $y=\tilde y_1+\varepsilon$, where $\varepsilon
>0$ is small, and use the initial conditions $G(\tilde
y_1+\varepsilon)$ and $G'(\tilde y_1+\varepsilon)$ from the series
expansions given in Proposition \ref{DiffSolutions}. In addition, we
use the numerical integration procedure to compute the integral to
scale the density $P$, by adding an additional ordinary differential
equation to (\ref{Diff}). 

\begin{figure}
  \centering
  \includegraphics[width=0.49\textwidth]{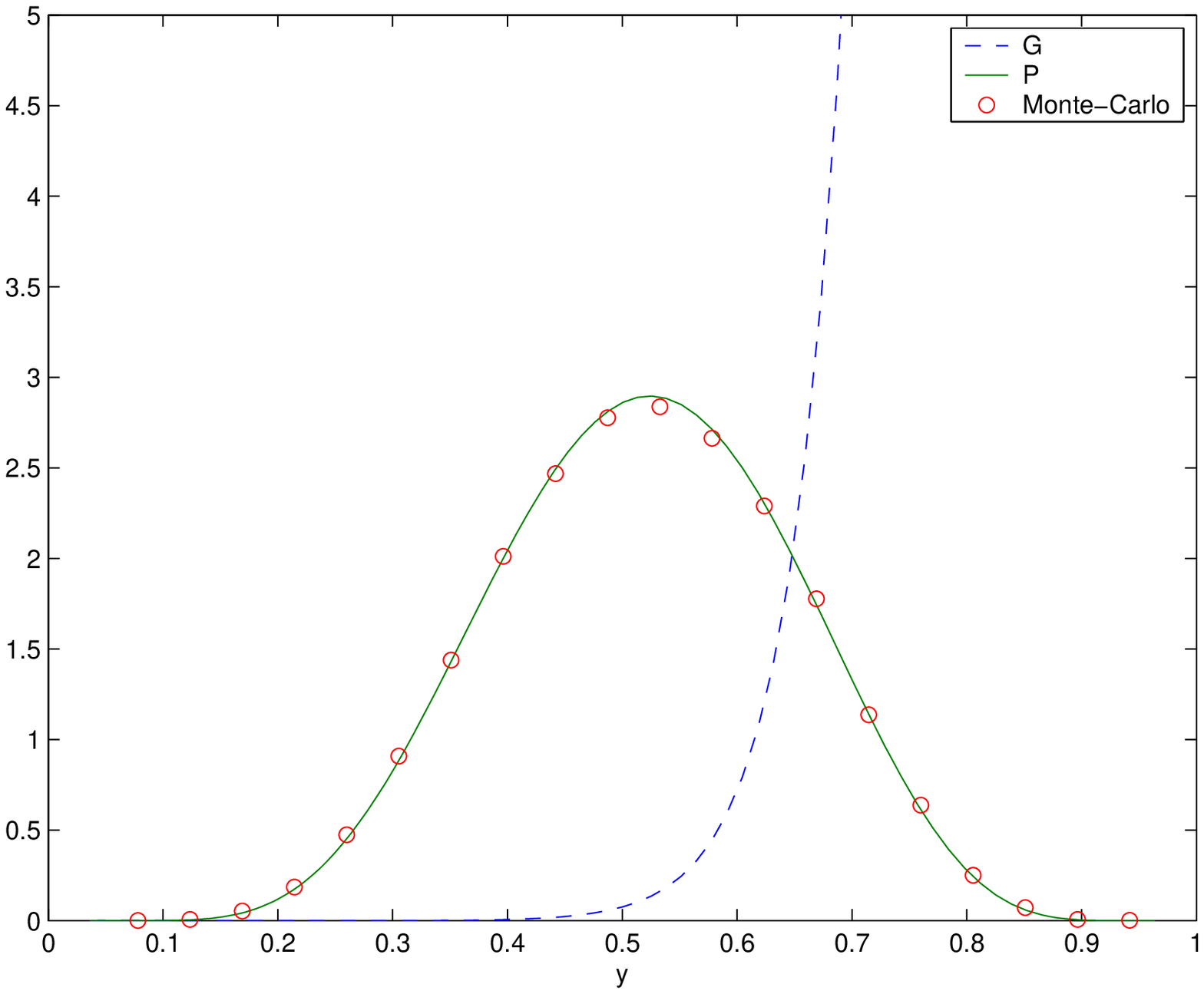}
  \includegraphics[width=0.49\textwidth]{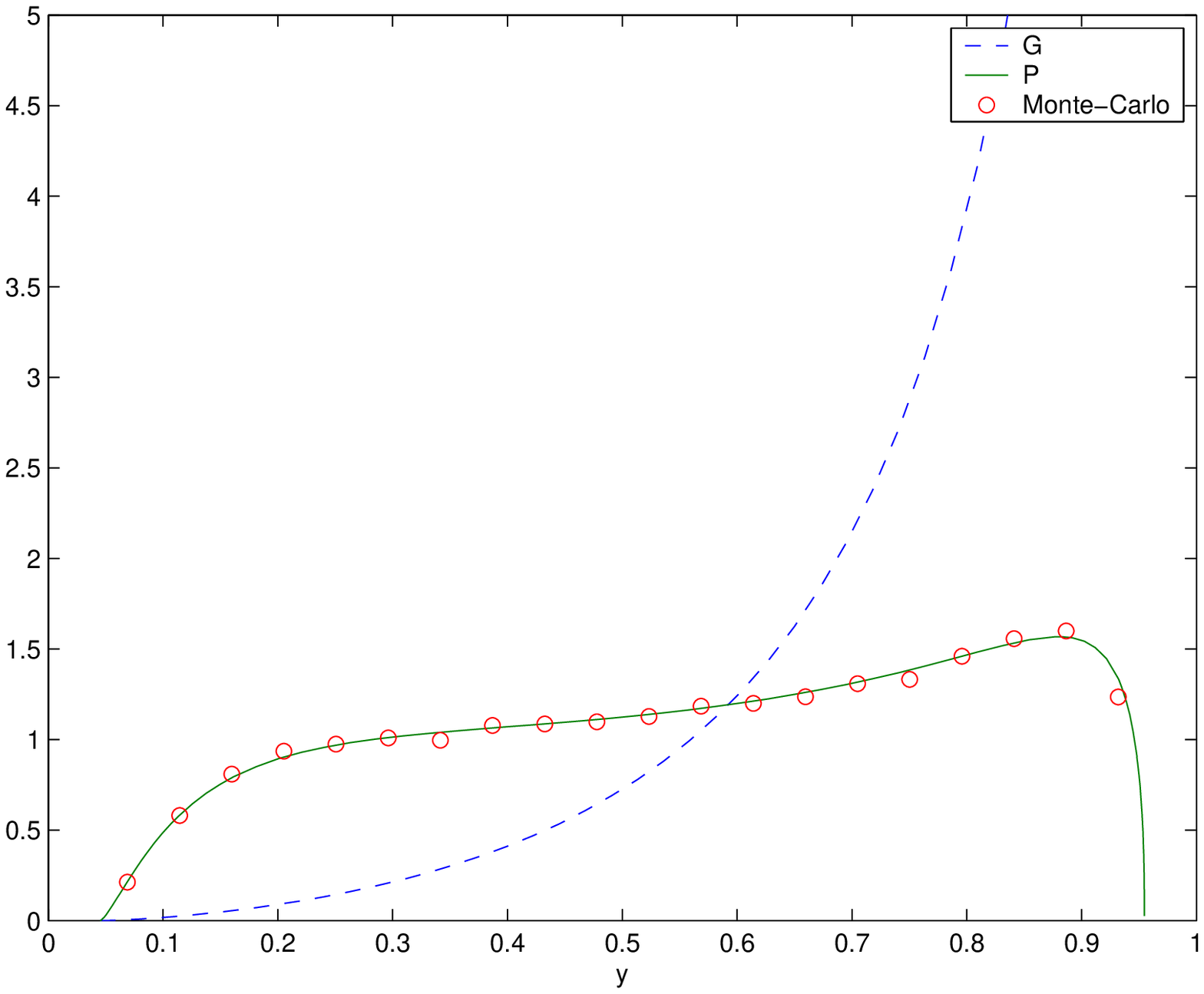}
  \caption{Density $P$ on the left when $\kappa=10$,
    $\triangle\gamma=1$, $k_0=0.4$, $k_1=0.05$, and on the right when
    $\kappa =1.5$, $\triangle\gamma=1$, $k_0=0.1$, $k_1=0.05$.}
  \label{fig12}
\end{figure}

\begin{figure}
  \centering
  \includegraphics[width=0.49\textwidth]{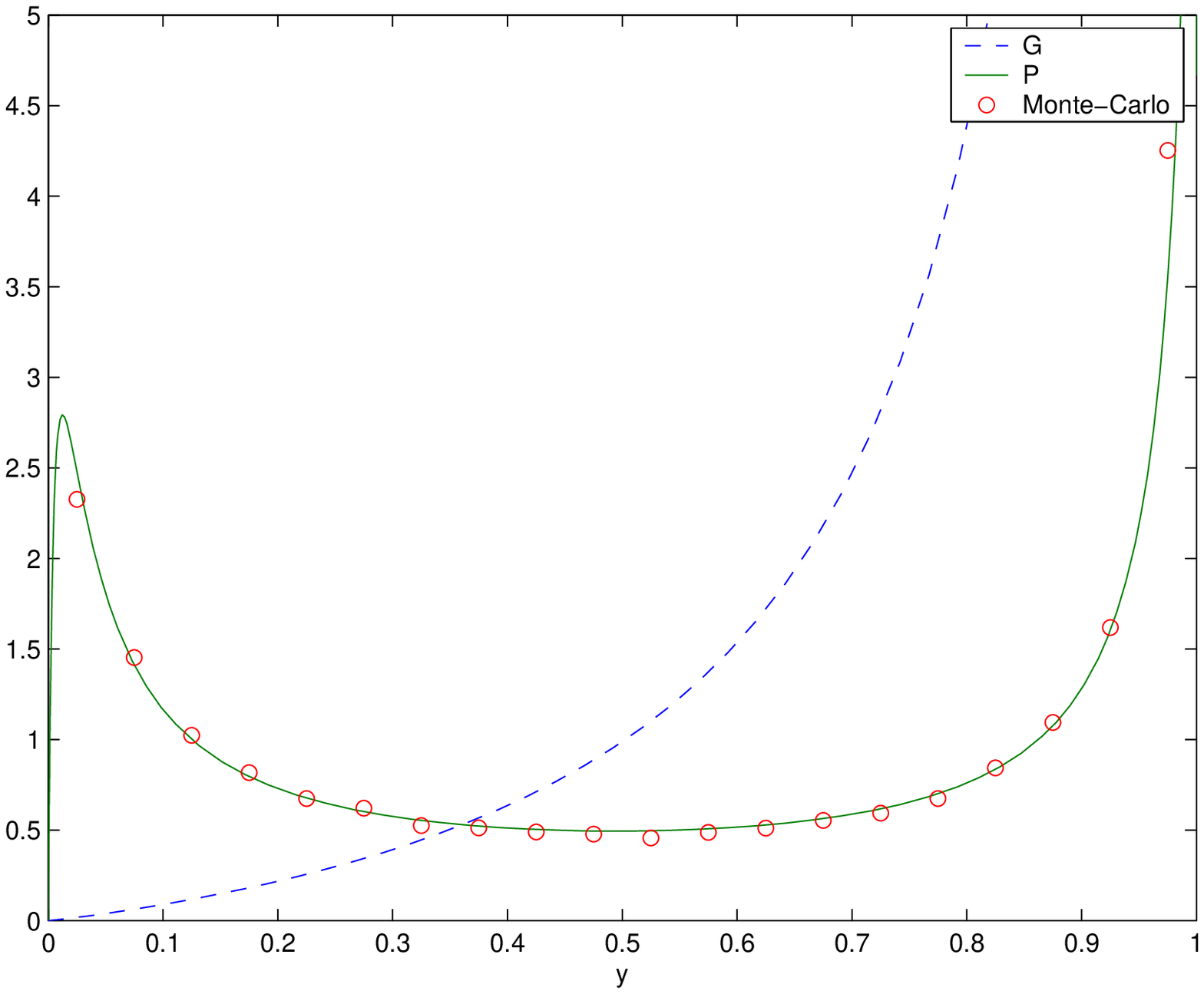}
  \includegraphics[width=0.49\textwidth]{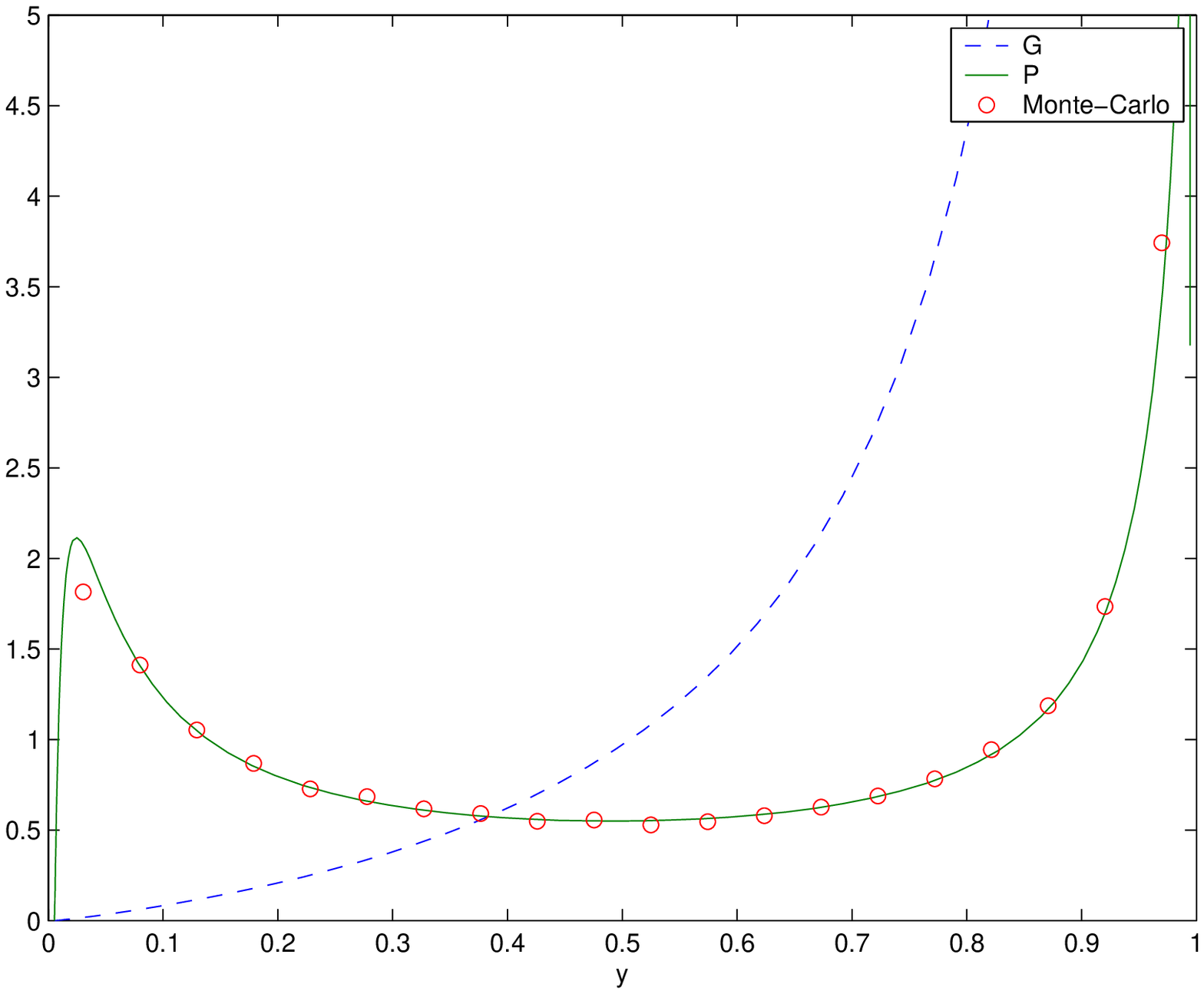}
  \caption{Density $P$ on the left when $\kappa =10$,
    $\triangle\gamma=9$, $k_0=0.1$,
    $k_1=0$. $Av(f_1)_{k_1=0}=0.553274111$, and on the right when
    $\kappa =10$, $\triangle\gamma=9$, $k_0=0.1$,
    $k_1=0.05$. $Av(f_1)_{k_1=0.05}=0.55672212$. Clearly the average
    fitness is larger when $k_1=0.05$ than when $k_1=0$.}
  \label{fig34}
\end{figure}

\begin{figure}
  \centering
  \includegraphics[width=0.49\textwidth]{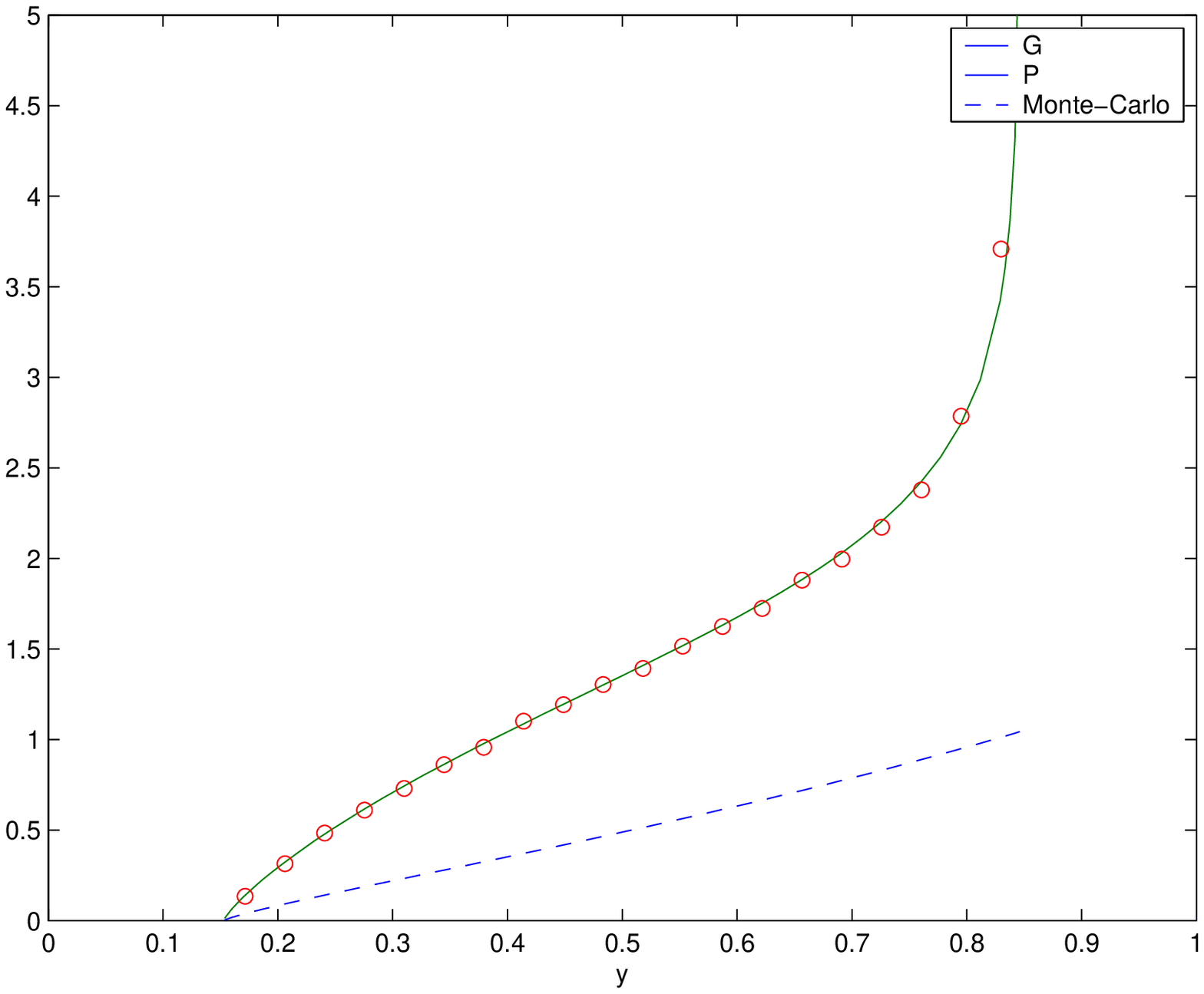}
  \caption{Density $P$ when $\kappa=5$,
    $\triangle\gamma=3$, $k_0=3$, $k_1=1$, a case where 
    $\alpha<1$.}
  \label{fig0}
\end{figure}

We show in Figures \ref{fig12} to \ref{fig0} the results obtained for
five different sets of parameters. 
In all the figures, we show the computed solution $G$ of the
differential equation (\ref{Diff}) in dashed, the computed density $P$
as a solid line, and the results of a Monte-Carlo simulation with
100'000 samples as circles. The density from the theory and the
Monte-Carlo simulations agree very well. It is interesting to see in
Figures \ref{fig12} and \ref{fig34} the variety of densities that can
be generated by this simple model. Figure \ref{fig34} contains a case
where increasing $k_1$ increases the overall fitness of the
population. Figure \ref{fig0} finally shows a case where $\alpha<1$.  We
note that the numerical integration out of the singularity can be
challenging. In particular, for the first case in Figure \ref{fig12},
the standard ode45 from Matlab needed very small absolute tolerances
to succeed with the integration for $\varepsilon<1e-2$. A more robust
method turned out to be DOPRI853, see \cite{Hairer}.

\pagebreak

\section{Appendix}

In this appendix we show for completeness the proof of Proposition
\ref{DiffSolutions} and describe a method how to solve the
differential equation (\ref{Diff}) (see also \cite{Jaenich},
pp.~317-321).  This equation is of the form
$$G''(y)+U(y)G'(y)+V(y)G(y)=0,$$
where the functions $U(y)$ and $V(y)$ are meromorphic in the complex plane
with four poles of order one at
$y_0<\tilde y_1:=1-y_1<y_1<\tilde y_0:=1-y_0$.
The solutions are therefore analytic in the open disc of radius
$(y_1-\tilde y_1)/2$ centered at $1/2$.
We look for real solutions in
the interval
$I=(\tilde y_1,y_1)$. In order to simplify calculations, we use
the variable transformation
\begin{equation} \label{transform}
      y=\tilde y_1+(y_1-\tilde y_1)z,\qquad z=\frac{y-\tilde 
y_1}{y_1-\tilde y_1}
\end{equation}
and set  $g(z):=G(y)$.
With this transformation, the differential equation (\ref{Diff}) becomes
\begin{equation}\label{newDiff}
g''(z)+u(z)g'(z)+v(z)g(z)=0,
\end{equation}
where
$u$ and $v$ have four poles of order one at the points $-b<0<1<1+b$ with
$b=(\tilde y_1-y_0)/(y_1-\tilde y_1)$:
$$u(z)=\frac{1-\alpha}z+\frac\alpha{z-1}-\frac\alpha{z+b}+\frac{\alpha 
+1}{z-(1+b)},\quad
v(z)=
\frac{\alpha(1+b)^2}{z(1-z)(z+b)(1+b-z)}.$$
We can therefore rewrite this equation as
\begin{equation} \label{genDiff}
     g''(z)+\frac{h(z)}zg'(z)+\frac{k(z)}{z^2}g(z)=0,
\end{equation}
where $h(z)$ and $k(z)$ are analytic in the disc of radius $\min\{1,b\}$
centered at 0:
$$h(z)=\sum_{n=0}^\infty \alpha_nz^n,\qquad k(z)=\sum_{n=0}^\infty
\beta_nz^n.$$
Multiplying the equation (\ref{genDiff}) by $z^2$ we get an
equivalent equation
which can be written as
\begin{equation} \label{opDiff}
     {\rm L}(g):=({\mu_z}^2{\rm D}^2+\mu_h\mu_z{\rm D}+\mu_k)(g)=0,
\end{equation}
where ${\rm D}$ denotes differentiation and $\mu_f$ multiplication
by a function $f(z)$.
Looking for solutions of the form
$$g(z)=z^\rho w(z),\qquad w(z)=1+\sum_{n=1}^\infty w_nz^n,$$
we may identify the function $g(z)$ with the infinite row
$[w]=[1,w_1,w_2,w_3,\ldots]$ and write (\ref{opDiff}) in matrix form:
\begin{equation} \label{linDiff}
     [{\rm L^\rho}][w]^{\rm T}=0.
\end{equation}
If we write ${\rm L}$
as
${\rm L}=(\mu_z{\rm D}+\mu_{h-1})\mu_z{\rm D}+\mu_k$, we get the lower
triangular matrix
$$[{\rm L^\rho}]=\left[\begin{array}{ccccc}
     \rho(\rho+\alpha_0-1)+\beta_0&0&\ldots\\
     \rho \alpha_1+\beta_1&(\rho+1)(\rho+\alpha_0)+\beta_0&0&\ldots\\
    \rho \alpha_2+\beta_2&(\rho+1)\alpha_1+
\beta_1&(\rho+2)(\rho+1+\alpha_0)+\beta_0&0&\ldots\\
     \vdots&\vdots&\vdots&\vdots&\ddots
     \end{array}\right].
$$
A solution $[w]=[1,w_1,w_2,\ldots]$ of the linear system (\ref{linDiff})
exists if and only if $L_{00}^\rho=0$. This is the so-called indicial
equation for $\rho$. From now on we shall no longer treat the general case
but only the case corresponding to our differential equation
(\ref{newDiff}). In this case $\alpha_0=1-\alpha$ and $\beta_0=0$. So the
indicial equation is
$\rho(\rho-\alpha)=0$
and yields the two characteristic exponents
$\rho_1=\alpha$ and $\rho_2=0$.
We shall write $L_{ij}^\nu$ instead of
$L_{ij}^{\rho_\nu}$.

For $\rho=\rho_1$, the solution
$[w^{(1)}]=[1,w_1^{(1)},w_2^{(1)},\ldots]$ may be calculated by the
recursion scheme
$$w_0^{(1)}=1,
\quad w_n^{(1)}=\frac{-1}{
L_{nn}^1}\left(\sum_{j=0}^{n-1}L_{nj}^1w_j^{(1)}
\right)\ {\rm for}\ n\ge1.$$
With these coefficients $w_n^{(1)}$, the function
$$g_1(z)=z^{\rho_1}\left(1+\sum_{n=1}^\infty w_n^{(1)}z^n\right)$$
is a solution of (\ref{newDiff}). From the general theory of linear
differential equations in the
complex plane it follows that $g_1$ is analytic in the disc of radius
$1/2$ centered
at $1/2$, but the power series for $w_1(z)$ might have a convergence
radius $0<\delta<1$.

If $\alpha$ is not an integer, another solution $g_2(z)$, linearly
independent of
$g_1(z)$, can be obtained in the same way from $\rho=\rho_2=0$. If, however,
$\alpha$ is an integer, the corresponding matrix has the entry ${\rm
L}_{nn}^2=0$
for $n=\alpha$, and we look in this case for a solution $g_2(z)$ of the form
$g_2(z)=1+\sum_{n\ge1} w_n^{(2)}z^n+Cg_1(z)\ln z$.
As $g_1$ is a solution, the terms in ${\rm L}(g_2)$
containing $\ln z$ cancel and the function
$w^{(2)}(z)=1+\sum_{n\ge1}w_n^{(2)}z^n$ must satisfy the equation
$${\rm L}(w^{(2)})=-C(2\mu_z{\rm D}+\mu_{h-1})(g_1).$$
Identifying $w^{(2)}(z)$ with the infinite row
$[w^{(2)}]=[1,w_1^{(2)},w_2^{(2)},\ldots]$,
we can write this in matrix form
\begin{equation} \label{linDiff2}
     [{\rm L^2}][w^{(2)}]^{\rm T}=-C[v_1,v_2,\ldots]^{\rm T}.
\end{equation}
For the right-hand side one checks easily that $v_j=0$ for
$j=0,\ldots,\alpha-1$ and
$v_\alpha=\alpha$. Therefore we can resolve the inhomogeneous linear
system (\ref{linDiff2}) in the following way:
\begin{enumerate}

     \item
       We determine $w_j^{(2)}$ for $j\le\alpha$ in the same way as
       $w_j^{(1)}$.

     \item
       We set $w_\alpha^{(2)}:=0$ and determine the constant $C$ by the equation
       $\sum_{j=0}^{\alpha-1} L_{\alpha,j}^{(2)}w_j{(2)}=-Cv_{\alpha}$.

     \item
       We determine the coefficients $w_n^{(2)}$ for $n>\alpha$ by the recursion
       formula
       $$w_n^{(2)}=\frac{-1}{
         L_{nn}^2}\left(Cv_n+\sum_{j=0}^{n-1}L_{nj}^2w_j^{(2)}
       \right)\ {\rm for}\ n\ge\alpha+1.$$

\end{enumerate}

We shall not go into further details, for example present concrete formulas
expressing the $v_n$ by the $w_n^{(1)}$, because we don't really need
the solution
$g_2$ of (\ref{genDiff}) in our case, as we have shown in the proof of Theorem
\ref{ergodic}.

Using the variable transformation (\ref{transform}) we get the
solutions $\tilde G_j(y)$ of
the original differential equation (\ref{Diff}), in particular
$$\tilde G_1(y)=(y_1-\tilde y_1)^\alpha g_1\left(\frac{y-\tilde
y_1}{y_1-\tilde y_1}\right)
=(y-\tilde y_1)^\alpha \tilde W_1(y)=(y-\tilde y_1)^\alpha\left(
1+\sum_{n=1}^\infty
     \frac{w_n^{(1)}}{(y_1-\tilde y_1)^n}\right).$$

     In order to find fundamental solutions near the singularity $y_1$, we can
apply the same method once more, but using the variable transformation
$$y=y_1-(y_1-\tilde y_1)z,\qquad z=\frac{y_1-y}{y_1-\tilde y_1}.$$ One easily
checks that in this case the indicial equation is
$\rho(\rho+\alpha-1)=0$
and that therefore the two characteristic exponents at $y_1$ are
$\rho'_1=1-\alpha\ {\rm and}\ \rho'_2=0.$ We obtain thus the second fundamental
system of solutions $G_1(y)$ and $G_2(y)$.



\noindent{\bf Acknowledgment}: The authors thank their collaborator and friend Gerhard Wanner for
many fruitful discussions concerning numerical solutions of differential
equations.





\end{document}